# A Novel Trust-Based DDoS Cyberattack Detection Model for Smart Business Environments

Oghenetejiri Okporokpo, Funminiyi Olajide, Nemitari Ajienka and Xiaoqi Ma

Department of Computer Science, Nottingham Trent University, Clifton Lane, Nottingham NG11 8NS

## ABSTRACT

*As the frequency and complexity of Distributed Denial-of-Service (DDoS) attacks continue to increase, the level of threats posed to Smart Internet of Things (SIoT) business environments have also increased. These environments generally have several interconnected SIoT systems and devices that are integral to daily operations, usually depending on cloud infrastructure and real-time data analytics, which require continuous availability and secure data exchange. Conventional detection mechanisms, while useful in static or traditional network environments, often are inadequate in responding to the needs of these dynamic and diverse SIoT networks. In this paper, we introduce a novel trust-based DDoS detection model tailored to meet the unique requirements of smart business environments. The proposed model incorporates a trust evaluation engine that continuously monitors node behaviour, calculating trust scores based on packet delivery ratio, response time, and anomaly detection. These trust metrics are then aggregated by a central trust-based repository that uses inherent trust values to identify traffic patterns indicative of DDoS attacks. By integrating both trust scores and central trust-based outputs, the trust calculation is enhanced, ensuring that threats are accurately identified and addressed in real-time. The model demonstrated a significant improvement in detection accuracy, and a low false-positive rate with enhanced scalability and adaptability under TCP SYN, Ping Flood, and UDP Flood attacks. The results show that a trust-based approach provides an effective, lightweight alternative for securing resource-constrained business IoT environments.*

## 1. INTRODUCTION

The deployment of smart internet of things (SIoT) in business environments typically involves a variety of interconnected hardware, such as RFID sensors, actuators, and microcontrollers. These systems enable the continuous monitoring of critical processes, remote access control of key business infrastructure, and help in decision-making processes such as real-time data analytics [1]. In industrial facilities, smart sensors often monitor machine health in a bid to avoid unplanned system downtime [2]. In logistics, GPS-enabled trackers assist in end-to-end visibility of shipments [3]. Retailers use SIoT systems to manage inventories, and in office environments, smart IoT devices are used to manage climate and lighting systems to enhance employee comfort and energy efficiency[4].

One of the main reasons IoT is adopted by most business operations is largely due to the ease of access and relatively reduced cost of these devices, as well as the increased demand for digital transformation [5]. However, as the IoT landscape continues to expand rapidly, there has been a substantial increase in cybersecurity challenges. Due to cost constraints, limited hardware

          49

capabilities, or insufficient memory capacity, SIoT devices are generally deployed with minimal security configurations [6]. As a result, these devices are highly prone to cyberattacks, particularly botnet attacks that exploit vulnerable nodes for large-scale distributed denial-of-service (DDoS) attacks [7].

The consequences of such attacks are extensive within smart business environments. DDoS attacks can disrupt critical business services, compromise sensitive data, and lead to significant financial losses. A single compromised node or device can act as an entry point for malware, allowing attackers to gain control over a network of devices [8].

Existing security mechanisms for SIoT devices are often inadequate in business settings due to their reliance on high resource availability [9]. Deep learning-based intrusion detection systems and edge-based security systems demand substantial processing power and memory, which are not feasible for the vast majority of smart IoT networks [10]. Moreover, complex security infrastructures involving multiple servers or hardware clusters introduce additional cost and maintenance overheads, which make them impractical for widespread deployment in enterprise business environments [11].

To address these limitations, there is a need for a lightweight, scalable, and autonomous security technique specifically designed for smart business environments. A solution that can detect and mitigate threats efficiently without imposing significant computational or administrative overheads [12]. This paper proposes such a solution, a lightweight trust-based security system for smart IoT business environments that aims to detect and contain DDoS attacks by monitoring network behaviour and identifying anomalous traffic patterns.

The technique is evaluated through an experimental setup. In the first phase, a SIoT device within a modelled business network is infected with malware and used to propagate the infection to other devices within the network, culminating in a coordinated DDoS attack. In the second phase, a trust-based attack detection mechanism is deployed to carry out the traffic analysis at the edge router to identify compromised nodes and devices. Trust scores are shared by the system with other nodes within the network, and the compromised nodes and devices are isolated in the network to prevent further propagation and ensure network resilience.

This study contributes to the field by providing a practical and resource-efficient technique for securing SIoT devices within business environments, which addresses both the technical constraints of smart IoT devices and the unique operational requirements of modern businesses. The key contributions of this paper are as follows.

- The approach incorporates Knowledge-based and observational-based trust computations, resulting in more efficient and effective trust aggregations in smart business environments.
- Central-based trust mechanism is deployed for updating trust and reputation, which retains the trust values of all the devices in the network allowing the network parameters to be adjusted as needed.
- The technique employs a hybrid trust-based approach for trust propagation, updating and formation. It classifies malicious nodes using knowledge, reputation, observational experience, and a central trust management system, resulting in improved identification and mitigation of security threats in smart business environments.

The remainder of this paper is organised as follows: Section 2 reviews related literature and current security approaches. In Section 3, we outline the methodology of the proposed technique.

Section 4 presents the proposed trust-based system in detail. In Section 5, we discuss the experimental results and evaluate the system performance. Finally, in Section 6, directions for future work are highlighted.

## 2. RELATED WORK

### 2.1. DDoS Threats in IoT and Smart Business Environments

The resource-constrained nature of SIoT devices creates vulnerabilities that are often exploited by cybercriminals. Most cybersecurity techniques, such as encryption, deep packet inspection, or frequent firmware updates, require huge computational power, memory, or hardened firmware for implementation, and many SIoT devices lack these. In many cases, devices operate with factory default credentials and minimal or no authentication, which enables attackers to gain access and adapt them into botnet agents for coordinated attacks. Smart business environments generally rely heavily on real-time data exchanges and cloud service communications, which increases the susceptibility of these networks to saturation and service exhaustion attacks.

In recent times, SIoT networks have been plagued with several DDoS cybersecurity challenges mainly due to the ubiquitous nature of these IoT devices and poor configurations of business edge routers [13]. In the last decade, the number and sophistication of security attacks on SIoT networks have increased significantly [14]. In most of the cyberattacks, the IoT devices in smart business environments are either taken over and used as bots for attacking a victim server, or they are the subject of these attacks, with cybercriminals using a variety of strategies to gain access and obtain sensitive information. A common example of this is the popular botnet called Mirai, which was launched in October 2016, several infected IoT devices were used to flood a victim's DNS server with lots of data [15-17].

Another IoT botnet, Gafgyt, was deployed to launch DDoS attacks targeting the victim servers [18-20]. The Gafgyt botnet, written in C and mainly infects Linux-based systems, is comprised of IoT devices such as DVR's, cameras and routers. Gafgyt infects new devices by leveraging Metasploit modules or scanning for open Telnet ports and then propagates different multi-vector DDoS attacks, including TCP flooding by exploiting TCP packet flags, keeping TCP connections open and flooding a specific TCP or UDP port with a large volume of data.

When a DDoS attack occurs, the disruption can halt automated business processes, disable accesscontrol systems, disrupt environmental monitoring systems, and compromise safety mechanisms, thereby leading to operational downtime and financial losses. Over the years, several mechanisms have been developed for DDoS detection and mitigation, broadly categorised into signature-based and anomaly-based techniques. Signature-based methods compare incoming traffic patterns with known attack signatures, offering high-speed detection but struggling with previously unseen threats. In contrast, anomaly-based methods analyse deviations from normal traffic patterns, providing the ability to detect novel attacks but frequently result in higher false-positive rates [21].

### 2.2. Common DDoS Attack Detection Techniques

Distributed Denial of Service (DDoS) attacks continue to pose a significant threat to modern businesses and SIoT networks, primarily due to their scalability, low resource requirements, and ability to exploit communication vulnerabilities. Researchers have explored a wide range of detection and mitigation strategies, each contributing unique perspectives but also facing recurring challenges regarding scalability, adaptability, and computational efficiency.

As indicated in the traditional techniques for DDoS detection, researchers have increasingly used rate-limiting and connection limits-based techniques to overcome these limitations. Rate-limiting techniques focus on detecting and mitigating DDoS attacks by restricting metrics, such as frequency of requests and packet rates within specific time frames [22]. On the other hand, connection limiting systems restrict the number of concurrent connections to prevent DDoS attacks [23]. However, many existing detection systems operate within isolated environments or lack scalability, which makes them impractical for deployment for SIoT networks. This restricts their practicality in smart business environments where density and traffic volatility are high. Additionally, they often do not integrate dynamic learning mechanisms that can adjust to everevolving cyber threats [24].

A paper by Javed et al. [25] presented a machine learning-based intrusion detection mechanism for smart homes. The proposed solution divides the IDS between a microcontroller-based smart thermostat (edge layer) and a cloud server (cloud layer). The edge layer handles initial detection and forwards data to the cloud, where classification is performed. However, a major limitation lies in computational overhead as the network size scales.

Kumar and Keshri [26] presented a Game Theory-based Adaptive Security (GT-AS) system for Distributed Denial of Service (DDoS) attack mitigation in IoT networks. The model is complemented by a Recurrent Bat (RB) framework for classifying nodes as either safe or malicious, for attack detection while managing energy consumption. Although a 47% increase in accuracy and error rate was achieved in the study compared to traditional methods, there is a lack of clarity into the comparative baseline models used.

Author Sanli [27] explored the security challenges associated with DDoS attacks in smart home environments. The approach incorporates buffer management and bitwise packet operations within an FPGA-based architecture for attack detection. The system demonstrates promising performance under various attack scenarios, with low resource overhead and response times. While the solution is tailored for smart home environments, the research primarily focused on hardware-level optimisation and lacks depth in evaluating adaptability to evolving attack patterns or diverse IoT environments.

Alam et al. [28] investigated methods for DDoS attack detection, highlighting how machine learning techniques, particularly neural networks and time-series forecasting, can significantly improve early DDoS attack detection and real-time grid monitoring. However, some key challenges with the techniques include data privacy concerns, scalability limitations, and difficulty in integrating with legacy systems.

Alemu et al. [29] highlighted the effectiveness of data communication networks as a key factor in the functionality of IoT networks. Their study focused on developing a detection mechanism tailored to MQTT-based DDoS attacks and evaluated the system against protocol-compliant attack scenarios. While the solution shows promise, its reliance on a centralised analysis module raises scalability and deployment challenges in highly dynamic vehicular environments.

Khaliq et al. [30] examined botnet propagation strategies in IoT networks, with a focus on how botnets are deployed to exploit the limited resources of IoT devices, including those in Wireless Sensor Networks (WSNs). A pandemic modelling approach was used to analyse how such botnets propagate in constrained environments. Common threats like the Mirai botnet were used in their study, which employed various machine learning and data mining techniques, including LSVM, neural networks, as well as decision trees to detect anomalous behaviour characteristic of DDoS attacks. Although the approach offers valuable insight into privacy-by-design strategies for IoT security, the centralised architecture presents resilience limitations.

Pillai et al. [31] in their study focused on implementing a software-defined networking (SDN) technique as a means of addressing the persistent threat posed by DDoS attacks on Cyber-Physical Production Systems (CPPSs) within Industry 4.0 frameworks. The technique utilizes programmability and dynamic traffic control for redirecting malicious internet traffic to a designated scrubbing centre, where they are subsequently filtered out prior to reaching the intended network or device. While this approach offers enhanced flexibility and responsiveness compared to static defences, it inherently depends on the resilience of the SDN control plane with very low fault tolerance.

Gupta et al. [32] applied an Optimised Edge-CNN (Convolutional Neural Network) model that uses a Cuckoo search algorithm for detecting DDoS attacks. Their model is tailored for use in resource-constrained environments. However, the evaluation methodology and comparison metrics were not clearly outlined, limiting the assessment of its real-world implementation and integration.

## 2.3. Trust-Based Security Models in SIoT Networks

Recent studies have further investigated the design of trust-based systems, which identifies, combines, and execute security decisions based on trust values and reputation with efficient aggregation functions to ensure data confidentiality and integrity in SIoT networks. These solutions are particularly relevant for devices with limited resource power [33-35]

Secure authentication and data management are critical components of SIoT security. Researchers have proposed a variety of methods to address the DDoS attacks that are prevalent in smart IoT networks. These approaches differ significantly, with each study emphasising distinct aspects of IoT network security [36-39].

Prathapchandran et al. [40] proposed a logistic-regression-based trust model for RPL-based IoT networks for detecting and isolating blackhole nodes. The technique uses direct, reputation, and experiential trust to aggregate a composite score used for malicious node classification. Although the results show an improved packet delivery ratio over other existing trust-based methods, the system is primarily designed to address only black hole attacks in IoT with limited adaptability to rapidly evolving attack scenarios.

A trust-based mechanism for mitigating hello-flood attacks in IoT networks was proposed by researchers Apurav and Saurabh [41]. The technique assigns trust values to nodes based on observed behaviour and isolates nodes with a low trust score. The system was evaluated in NS-2 using metrics including throughput, packet loss, energy usage, and delay. While the approach demonstrates improved performance in detecting hello-flood attacks, the trust model lacks consideration of lightweight trust computation suitable for constrained devices. Also, the system appears limited to single-attack scenarios with no adaptation for increased node volumes or evolving cyberattack patterns.

Wei et al. [42] presented a trust-based scheme for enhancing security in Mobile Ad Hoc Networks. Bayesian inference is used for computing trust values from an observer node, whilst the indirect observation is obtained from neighbour nodes. A key advantage of this approach is the rapid detection of node misbehaviours such as dropping or modifying packets. However, as the network size increases, the performance of the overall system degrades.

Based on our analysis of trust-based techniques and their implementations, we have identified several gaps in existing detection systems, including dependence on specific network topologies, an inability to detect various DDoS attack vectors, and dependence on outdated or inaccurate data. These highlight the need for a comprehensive model that includes all critical elements of security in smart business environments. Consequently, this research focuses on a proposed trust-based system designed to enhance the security of SIoT business environments. A trust-based management technique that offers a structured approach to building and maintaining trust within a network. By integrating trust as a fundamental element, the proposed model delivers an adaptive, lightweight solution that strengthens the overall security of smart business IoT environments.

## 3. METHODOLOGY

In this section, we outline the methodology deployed by the proposed system for detecting and mitigating DDoS attacks within Smart Business environments. The methodology is underpinned mainly by a trust management system modelled to have each node in the network maintain a trust score of other nodes based on their behaviour. The trust scores are computed continuously based on the ongoing interaction between nodes. When suspicious, malicious behaviour is detected, such as in the case of a DDoS attack or a node behaving selfishly, trust scores reduce, the system identifies the specific node and necessary actions to determine if it is a genuine attack are taken.

### 3.1. Network Architecture

Trust-based systems are largely made up of three unique features. Resilient nodes that support communication, nodes which can collate protocols used to compute and disseminate information pertaining to current and future interactions, as well as a real-time feedback mechanism to aid the decision-making process of trust formation within the network. Figure 1 below shows a highlevel overview of the proposed Trust-based system.

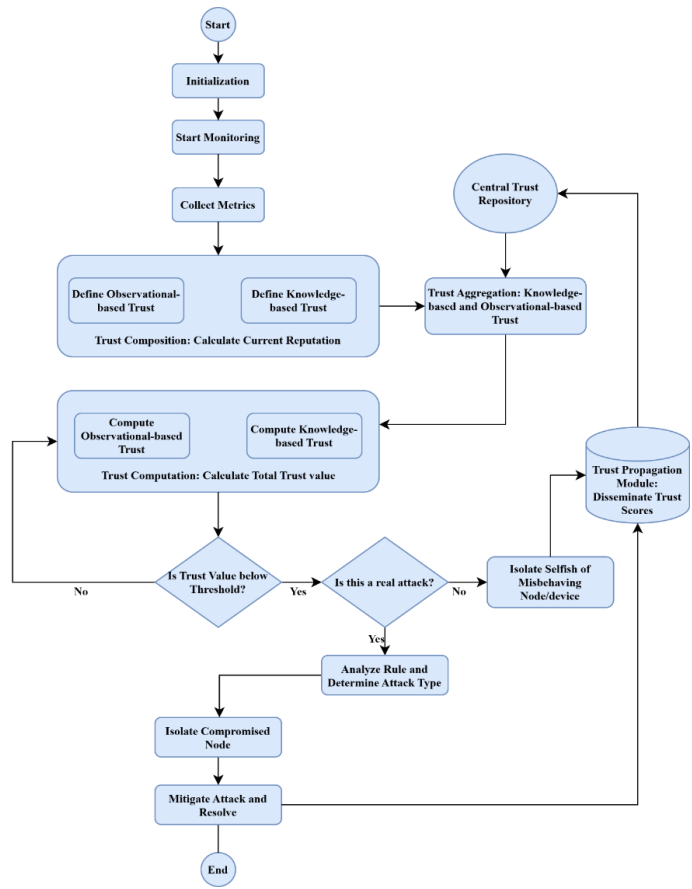

Figure 1. High-Level Overview of Proposed Trust-based System.

Evaluating the degree of security of a SIoT network is crucial for any setup within a smart business environment. We have compiled a comprehensive set of security parameters crucial for gauging security within a SIoT network environment. These parameters formulate the core of our trust model, resulting in a trust value as an outcome. This trust value offers a comprehensive perspective of the overall composition of the SIoT network security and can be broken down into several security properties based on these factors, expressed as a trust vector.

## 3.2. Hybrid Trust Definition

Trust is the degree of certainty that a node within the network is likely to perform based on predetermined protocols. This is measured by the degree of expectation of a node x on a peer node y to carry out a function, such as reliably forwarding packets without malicious intent [26]. In the proposed system, trust is defined based on a combination of knowledge-based and observational trust factors.

Packet Delivery Ratio (PDR): This is a measure of the degree of reliability of node y. The percentage of successfully delivered packets with respect to the total number of transmitted packets. A high PDR suggests that the node can be trusted to forward packets consistently, whereas a low PDR may indicate that the dropping or mishandling of packets by a node is deliberate and indicative of malicious behaviour or unintentional, which indicates a selfish node attempting to conserve resources.

$$PDR_{x,y}(t) = \frac{\text{Packets Delivered by node y}}{\text{Total Packets Sent to node y}} \qquad (1)$$

Anomaly Detection (AD): Anomaly detection is determined by sudden increases in the network traffic, which is usually the case when a node is the target of a DDoS attack. The system constantly inspects the network traffic, and the trust score decreases if an anomaly is detected.

$$AD_{x,y}(t) = \begin{cases} 1 & \text{if no anomaly is detected,} \\ 0 & \text{if an anomaly is detected.} \end{cases} \qquad (2)$$

Response Time (RT): This refers to how long a node takes to reply to communication requests from other nodes. When a node consistently has high response times, either because it is slow, unresponsive or overloaded, this could indicate that the node is either under attack and has been compromised.

$$RT_{x,y}(t) \frac{1}{\text{Observed Response Time of node y}} \qquad (3)$$

## 3.3. Trust Calculation

Trust is measured based on the degree of belief between communicating nodes based on mutually predefined parameters. The key parameters used in evaluating Trust in our proposed solution are packet delivery ratio, response time, and anomaly detection. The hybrid-based approach defines the knowledge-based trust and observational based trust as below.

**Knowledge-Based Trust:** Derived from predefined parameters, including historical node data, reputation, firmware updates, and security certifications. These factors provide a baseline trust level based on prior knowledge of the node's capabilities and expected behaviour.

**Observational-Based Trust:** Determined through real-time monitoring and analysis of node behaviour, such as packet delivery ratio, response times, and anomaly detection scores. Trust $T_{(x,y)}(t)$ between nodes x and y at any time t is computed based on a weighted sum of the three parameters: Packet Delivery Ratio, Anomaly Detection, and Response Time. The trust calculation formula is:

$$T_{x,y}(t) = \gamma \times KBT_{x,y} + \varepsilon \times OBT_{x,y}(t) \quad (4)$$

Where:

$T_{x,y}(t)$ is the overall trust value for node y as evaluated by node x at time t. $KBT_{x,y}$: Knowledge-based Trust score (Static or slowly updated).

$OBT_{x,y}(t)$: Observational-based Trust score (Dynamically updated).

**γ** and **ε**, are the weights for knowledge-based and observational-based trust, respectively, with **γ** + **ε** = 1, determined based on the specific requirements of the network.

**Calculation of Knowledge-Based Trust (KBT)**

The KBT score can be computed as a weighted average of the following parameters:

$$KBT_{x,y} = \alpha \times R_{x,y} + \beta \times S_{x,y} + \emptyset \times M_{x,y} \quad (5)$$

$R_{x,y}$ is the reputation of node y based on historical data or reviews from other trusted entities.
$S_{x,y}$ is the Security certificate score (that is, compliance with known standards).
$M_{x,y}$ is the main credibility score based on known vulnerabilities or trust level of the network.
**α**, **β**, and **∅** are the weights for these factors determined by network priorities.

**Calculation of Observational-Based Trust (OBT)**

The OBT score can be computed as a weighted average of the following parameters:

$$OBT_{x,y}(t) = \delta \times PDR_{x,y}(t) + \theta \times AD_{x,y}(t) + \mu \times RT_{x,y}(t) \quad (6)$$

Where:

**$OBT_{x,y}(t)$** is the observational-based trust value between nodes x and y at time t.
**$PDR_{x,y}(t)$** is the packet delivery ratio between node x and node y. That is the ratio of successful packet deliveries.
**$AD_{x,y}(t)$** is the anomaly detection score, indicating if node y's behaviour is deemed suspicious.
**$RT_{x,y}(t)$** is the response time of node y as observed by node x. That is the delay in responses from node y.
δ, θ, μ are the weights for each factor, with δ + θ + μ = 1, determined based on the specific requirements of the network. For example, if packet delivery is prioritised, δ would be larger.

## 3.4. Dynamic Trust Update Mechanism

Trust is a dynamic attribute that changes as node interactions occur over time. The system continuously evaluates the behaviour of each node, with trust scores updated at regular intervals using recent observational data. This adaptive design enables responsiveness to varied network conditions and evolving malicious attacks. The trust update process operates as follows:

$$T_{x,y}(t+1) = (1 - \omega) \times T_{x,y}(t) + \omega \times (\gamma \times KBT_{x,y} + \varepsilon \times OBT_{x,y}(t + 1)) \qquad (7)$$

Where $\omega$ is the update rate.

This dynamic update mechanism ensures responsiveness to the evolving behaviour of nodes. The parameter $\omega$ acts as a learning rate, balancing the influence of recent observations against historical trust. A high $\omega$ value makes the trust system more reactive to recent interactions, beneficial in rapidly changing environments. Conversely, a lower $\omega$ value provides more stability, avoiding drastic trust changes due to temporary anomalies.

Additionally, the trust update process incorporates the following features:

Decay Function for Inactivity: If a node does not interact for a defined time threshold, its trust value is reduced incrementally to reflect uncertainty:

$$T_{x,y}(t+1) = T_{x,y}(t) \times e^{-\lambda (t - t_{last})} \qquad (8)$$

where $\lambda$ is the decay rate and $t_{last}$ is the time of the last interaction.

Sliding Window Evaluation: Trust is computed over a moving time window $W$, aggregating observations to smooth short-term fluctuations.

Weighted Averaging of Observations: The model can apply recency-based weights to different observations within the window, giving more importance to the most recent behaviours. This multi-faceted update mechanism increases trust accuracy, supports adaptability to dynamic network conditions, and enhances the resilience of the smart network against security threats.

## 3.5. Trust Propagation in the Network

To accelerate trust dissemination and identify malicious nodes more efficiently, the model allows continuous trust propagation among nodes. When a node k lacks sufficient direct interaction data with node y, it may infer trust based on recommendations from a trusted intermediary node x.

This two-hop propagation is defined as:

$$T_{k,y}(t) = \frac{\gamma (KBT_{k,x} + KBT_{x,y})}{2} + \frac{\varepsilon (OBT_{k,x}(t) + OBT_{x,y}(t))}{2} \qquad (9)$$

This mechanism allows trust values to be inferred where direct observations are limited. Trust propagation assumes a transitive trust relationship but mitigates potential manipulation by averaging and by weighting knowledge and observational inputs separately.

## 4. THRESHOLD-BASED DETECTION SYSTEM

The threshold-based detection mechanism is a critical component of our hybrid trust framework, enabling the system to respond to suspicious or malicious activity by monitoring the trust levels of each node. When a node's trust score falls below the predefined threshold, $T_{thresh}$ it triggers a set of security actions aimed at mitigating potential threats.

- Monitor Trust Values: Maintains continuous observation of trust values within the network.
- Detect Malicious Nodes: Flags a node as potentially malicious or compromised when its trust score falls below the defined threshold ($T_{thresh}$), marking it for further review.
- Isolate Suspicious Nodes: Once a node is flagged, its ability to communicate with other nodes is temporarily limited until further review is carried out and the node is cleared.

**Trust Threshold Definition**: The threshold $T_{thresh}$ serves as a security benchmark for acceptable node behaviour. It is determined empirically based on network sensitivity, application-specific tolerance for risk, and the distribution of normal trust values. This value ranges between 0.3 and 0.5 on a normalized trust scale (0,1).

$$\text{If } T_{x,y}(t) < T_{thresh} \text{ then node y is flagged as suspicious.} \quad (10)$$

This condition is evaluated periodically and forms the basis for initiating countermeasures.
**Multi-Tier Detection Strategy:** To reduce false positives and provide nuanced responses, a multi-tier detection strategy is employed:

$$\text{Alert Zone } (T_{Alert}(t) \leq T_{x,y}(t) < T_{thresh}) \quad (11)$$

This triggers the node to be placed under enhanced observation. Additional metrics are collected at a higher frequency, and the node's trust evolution is reviewed for sudden changes.

$$\text{Isolation Zone } (T_{x,y}(t) < T_{thresh}) \quad (12)$$

The node is isolated from critical network paths and is either rate-limited, sandboxed, or restricted in its interactions with other nodes.

**Recovery Mechanism:** If the node recovers and its trust score remains above $T_{thresh}$ for a sustained period, it is gradually reintegrated into normal operation. This stratified approach enables adaptive security responses that balance vigilance and flexibility.

**Adaptive Threshold Tuning:** The threshold is not static; it can be dynamically adjusted based on environmental factors such as network load. During high congestion, the threshold may be lowered slightly to avoid over-flagging nodes under stress. It could also be dynamically adjusted based on device criticality. For high-value devices such as a router, a higher threshold is enforced.

**Anomaly Correlation:** If multiple correlated anomalies are detected, the effective threshold may be temporarily increased for suspicious nodes.

$$(T'_{thresh} = T_{thresh} + F_{event}) \quad (13)$$

Where $F_{event}$ is an event-sensitive adjustment factor.

**Integration with Trust Decay and Propagation:** Threshold evaluation is interdependent with trust decay and propagation mechanisms. Nodes with decaying trust due to inactivity may unjustly fall below the threshold. To counter this, threshold checks are only performed when sufficient recent interaction data exists. Indirect trust values obtained via propagation are assigned lower confidence when used for threshold-based decisions. This ensures that trust degradation due to transitive influence does not prematurely trigger defences.

**Response Policy:** Once a node is flagged as suspicious, neighbouring nodes are notified. Routing protocols bypass the node, and its network access privileges are downgraded. The event is logged for further analysis, including time, trust score, and anomaly evidence. This flexible technique is essential for proactively mitigating the impact of DDoS attacks.

**DDoS Attack Detection Algorithm**

Our hybrid trust model is tailored to effectively detect DDoS attacks. Target nodes that display abnormal behaviour are detected through the observational-based metrics. The algorithm used for detection is as follows:

Step 1: Initialise Trust Values:
Assign an initial neutral trust value for every peer device:
$T_{x,y}(0) = 0.5$
Step 2: Begin Continuous Monitoring
Monitor neighbour behaviour at regular intervals, collecting metrics:
$PDR_{x,y}(t), AD_{x,y}(t), RT_{x,y}(t)$
*Step 3: Calculate Observational Trust (OBT)*
Apply Equation (6) to calculate the dynamic behavioural trust contribution. *Step 4: Combine Observational Trust (OBT) with Knowledge-Based Trust (KBT)* Combine KBT and OBT using the hybrid trust Equation (4).
*Step 5: Update Trust Score*
Smooth trust updates with Equation (7) to avoid abrupt changes.
*Step 6: Compare Trust to Threshold*
If $T_{x,y}(t) < T_{thresh}$) then mark node y as suspicious
*Step 7: Trigger Mitigation Response*
Quarantine or isolate the suspicious node from further communication.
*Step 8: Continue Monitoring Suspicious Nodes*
Maintain observation and periodically re-evaluate trust score.
*Step 9: Restore or Permanently Block*

If trust improves and remains stable above $T_{thresh}$, reinstate node. Otherwise, permanently isolate.

## 5. RESULTS AND ANALYSIS

To evaluate the performance of the proposed model, an experiment was carried out using SIoT devices. To implement the trust-based detection module, we leverage the IoT device components with trust assessment functionality. Each device cumulates the trust score of its neighbour devices based on their observed actions (packet delivery, response time, and anomaly detection). The performance metrics collated include:

- Malicious Attack Detection Rate: A measure of the percentage of nodes accurately flagged as malicious by the system.

International Journal of Network Security & Its Applications (IJNSA) Vol.17, No.5/6, November 2025

- False Positive Rate: A measure of the percentage of benign nodes erroneously marked as malicious.
- Latency: A measure of the average duration for accurately detecting and mitigating an attack.
- Network Throughput: This is a measure of the total amount of data successfully transmitted within the network, demonstrating the effects of an attack on the overall network performance.

## 5.1. Experimental Environment

The hybrid trust-based DDoS detection model was evaluated under a combination of real SIoT traffic and scaled OMNeT++ simulation against TCP SYN flood, ICMP ping flood and UDP flood attacks. The physical and virtual components of the experimental environment are depicted in Figure 2 and summarised below:

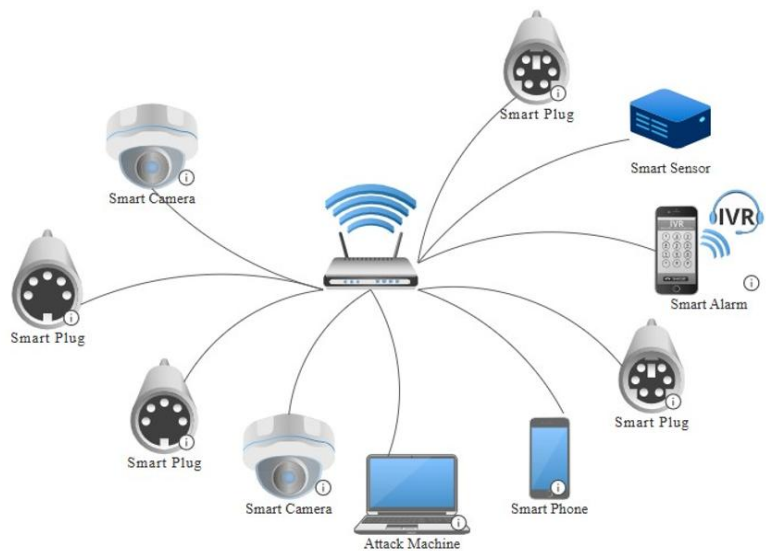

Figure 2. Topology of the network connection used in the test environment.

Smart IoT Devices used are:

- 2 × Smart Cameras (1080p IP cameras with RTSP streaming)
- 4 × Smart Plugs (energy-monitoring capable)
- 1 × Smart Lamp (Wi-Fi enabled, colour adjustable)
- 1 × Smart Door Sensor
- 1 × Smart Alarm (motion sensor + siren)

All devices communicated via standard home IoT protocols (MQTT, HTTP, HTTPS, and proprietary vendor APIs). They were connected to the local network with individual link speeds limited to 100 Mbps to reflect typical IoT hardware constraints.

Router: A Sagemcom F3896LG router was used to simulate a business-grade router with support for VLAN segmentation, static routing, and deep packet inspection (disabled to isolate trust-based detection performance).

Monitoring and Analysis System:

- A dedicated laptop running Kali Linux was connected to the router at 1 Gbps. It hosted:
- Wireshark for passive traffic capture and analysis.
- Custom Python scripts for real-time trust score computation based on observed device behaviour.

Mobile Device: A smartphone running the smart home applications was connected via Wi-Fi (256 Mbps average link speed). It received push notifications for alerts and provided a user interface for reviewing device status and manual overrides.

### 5.2. Trust Calculation Implementation

Each IoT device was assigned a baseline trust score. The knowledge-based trust was static. Observation-based trust was computed dynamically using metrics extracted from live traffic. Trust scores were updated at 10-second intervals and decayed over time in the absence of communication, allowing for gradual recovery or continued deterioration based on behavioural consistency. To ensure efficiency and proper configuration, the weighting coefficients, trust threshold and initial parameter selection are determined empirically through controlled iterative simulation experiments, which allow for reproducible traffic and attack conditions. Sensitivity analysis shown in Figures 3-6, was used to identify parameter sets that delivered high detection accuracy with low false positive rates. These parameters were then deployed on the modelled SIoT testbed. Parameters were retained only if they demonstrated consistent behaviour across both environments, ensuring that trust calculations remained consistent, responsive, and improved the overall trust computation performance.

Sensitivity analysis on a weight scale (0.10–0.90) revealed that detection accuracy peaked and the false positive rate reached its minimum at parameter configurations adopted in the final model.

**KBT/OBT**: $T = \gamma \times KBT_{x,y} + \varepsilon \times OBT_{x,y}(t)$ with $\gamma = 0.35$, $\varepsilon = 0.65$
$OBT$: $\delta = 0.45$ (PDR), $\theta = 0.35$ (AD), $\mu = 0.20$ (RT)
Update rate: $\omega = 0.4$
Inactive decay: $\lambda = 0.015 s^{-1}$

Thresholds: default $T_{thresh} = 0.40$ Alert band $T_{Alert} = 0.40\text{-}0.45$
Mitigation policy: Two consecutive violations $< T_{thresh}$ = Isolate, auto recovery after trust $> T_{thresh}$ for 30s.

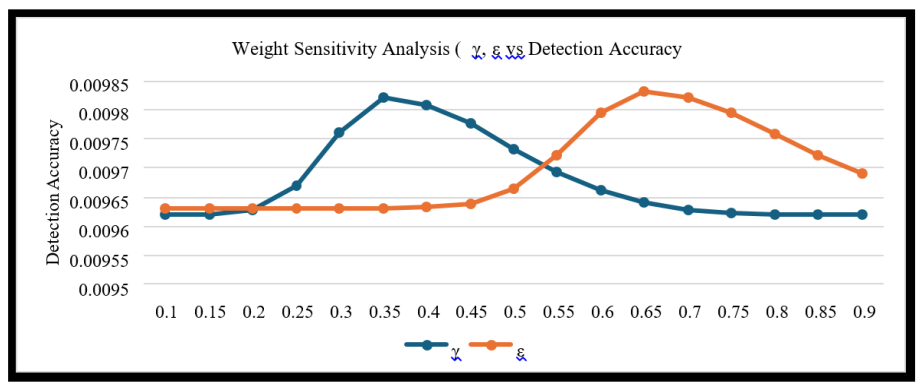

Figure 3. Sensitivity Analysis for Weighted Values (γ and ε).

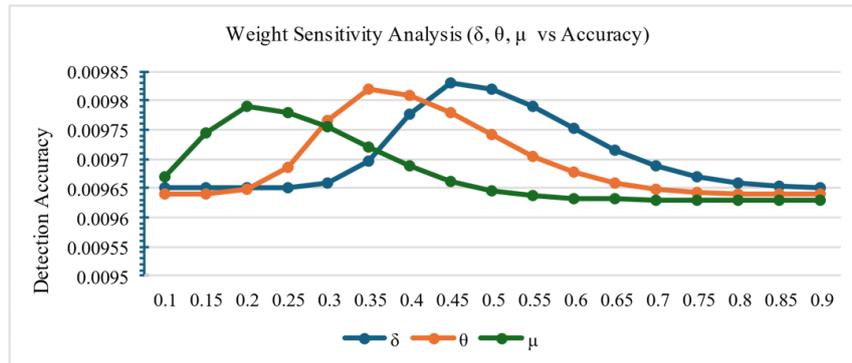

Figure 4. Sensitivity Analysis for Weighted Values (δ, θ and μ).

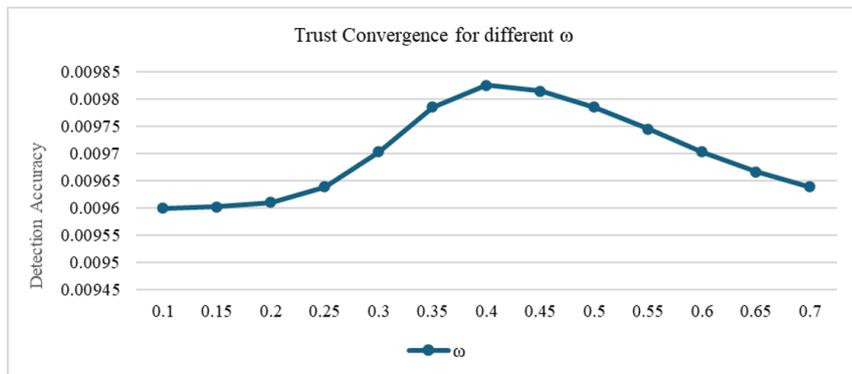

Figure 5. Sensitivity Analysis for Weighted Values (δ, θ and μ).

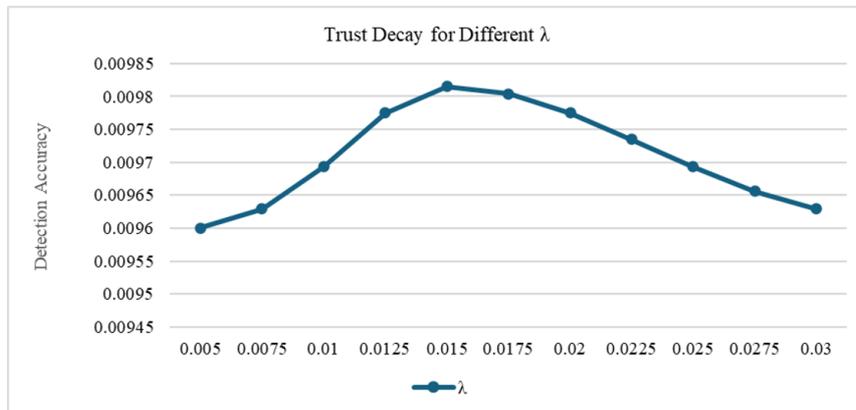

Figure 6. Sensitivity Analysis for Weighted Values (δ, θ and μ).

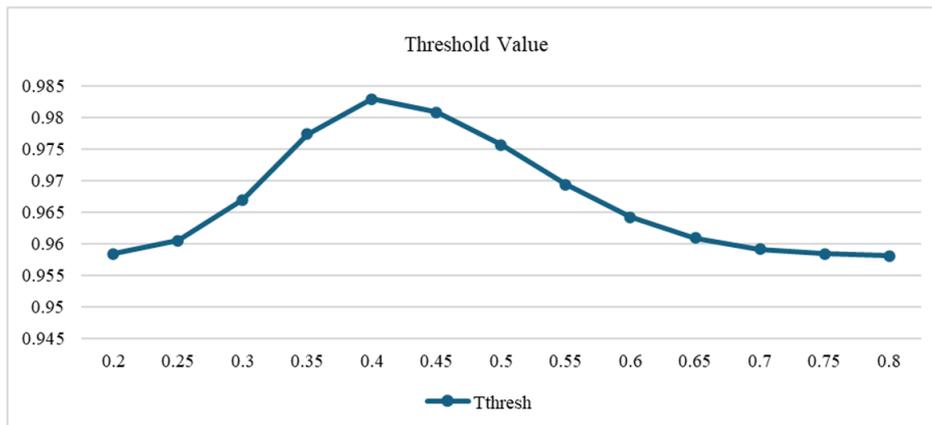

Figure 7. Sensitivity Analysis for Weighted Values (δ, θ and μ).

## 5.3. Attack Scenarios

### 5.3.1. Scenario 1

An attacker machine running Kali Linux 2025, hosted on a VirtualBox virtual machine, was deployed. The machine was configured to launch volumetric network attacks aimed at degrading the trustworthiness of targeted devices. Two specific flooding attack scenarios were executed during the experiments:

TCP SYN Flood (Attack 1): A high-rate SYN flood targeting multiple SIoT devices. This attack focused on depleting the connection queues consistent with volumetric attacks.

Ping Flood (Attack-2): A continuous stream of ICMP Echo Requests was directed toward randomly selected device ports and IPs. This attack focused on saturating the device processing capacity and network bandwidth, causing abnormal spikes and packet handling delays.

UDP Flood (Attack 3):

We generated continuous UDP traffic toward random device ports, mimicking a brute-force approach to overwhelm bandwidth and trigger abnormal device responses.

Each attack ran 10 minutes inside a 30-minute scenario; the benign background included camera streaming, plug telemetry, and sporadic control messages. Trust score variations, system responses, and alerting accuracy were analysed during and after the attack intervals. All experiments were logged using Wireshark and the custom trust engine. The environment was reset to a baseline state before each trial to ensure consistency and repeatability.

### 5.3.2. Scenario 2: Scalability & Lightweight Trust Based Solution

Our hybrid model introduces minimal computational and communication overhead, making it suitable for low-power smart devices. Wireshark PCAPs were exported as CSV and replayed using INET's traffic replayer. Trust logic was ported as a C++ module, devices and link rates mirrored the lab topology, then scaled. Scales tested: 8, 20, 30, 50 devices with identical per-class traffic profiles and a constant attacker traffic. The hybrid trust-based model maintains high detection accuracy and low overhead even as the number of devices and interactions increases. The simulation parameters are shown in Table 1.

Table 1. Common Simulation Parameters

| Simulation environment | Values |
| --- | --- |
| **Simulator** | OMNET++ v 6.0.2 |
| **Platform** | Windows 11 |
| **Number of Nodes** | 8 - 50 |
| **Time Interval** | 100-1000s |
| **Topology** | 800m X 600m |
| **Communication Range** | 50m |
| **Default Trust Value** | 0.5 |
| **Trust Threshold Value** | Data Link |
| **Malicious Penalty** | 0.2 |
| **Update Rate** | 0.4 |
| **Threshold Default** | 0.4 |

## 5.4. Results

Malicious Attack Detection Rate: We calculated the percentage of malicious nodes correctly flagged by the system. The hybrid trust-based system consistently achieved a higher detection rate, compared to GPSVM [43], SFaDMT [44], and DPLPLN [45], it detected attacks more quickly and more reliably due to its combined knowledge-based and observational trust mechanisms.

Table 2 Detection Rate

| Attack Type | Proposed System | GPSVM | SFaDMT | DPLPLN |
| --- | --- | --- | --- | --- |
| TCP SYN Flood | 99.1% | 95.01% | 97% | 91.5% |
| Ping Flood | 98.3% | 94.51% | 97% | 89.2% |
| UDP Flood | 98.8% | 93.6% | 96.4% | 98% |

Figure 8 below illustrates the comparison of our system with other systems and demonstrates that a higher detection rate is obtained by our system.

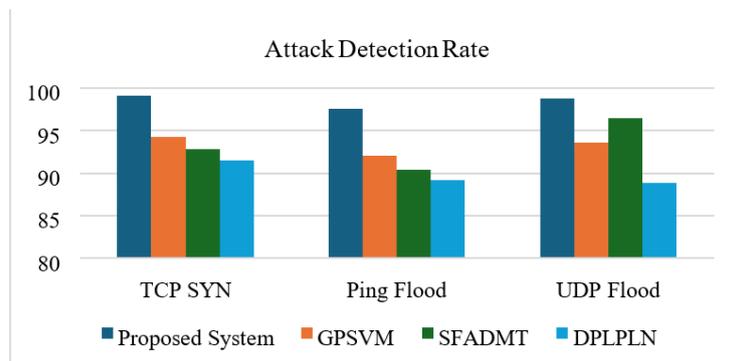

Figure 8. Malicious Attack Detection Rate

False Positive Rate (FPR): We measured the percentage of benign nodes incorrectly flagged as malicious. A lower FPR indicates greater reliability and reduces unnecessary node isolation.

Table 3 False Positive Rate

| Attack Type | Proposed System | GPSVM | SFaDMT | DPLPLN |
|---|---|---|---|---|
| TCP SYN Flood | 3.4% | 11.96% | 7.1% | 8.4% |
| Ping Flood | 2.8% | 7.88% | 7.9% | 8.8% |
| UDP Flood | 3.0% | 8.22% | 7.4% | 9.1% |

Our hybrid model demonstrates the lowest FPR across all evaluated attack scenarios. The integration of KBT prevents false alarms caused by temporary network fluctuations, while OBT dynamically responds promptly to sustained anomalies. This balance reduces misclassification and prevents legitimate nodes from being unintentionally isolated.

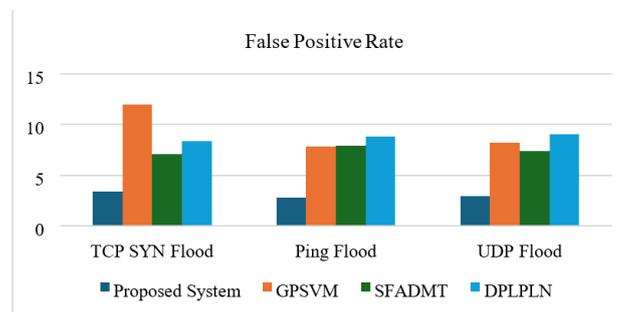

Figure 9. False Positive Rate

Latency: Here, we measured the time from the start of an attack to when the system detected and initiated mitigation. A shorter detection time indicates a more responsive security system that limits the attack impact more effectively.

Table 4. Latency

| Attack Type | Proposed System | GPSVM | SFaDMT | DPLPLN |
|---|---|---|---|---|
| TCP SYN Flood | 3.4ms | 6.2ms | 8.46ms | 8.4ms |
| Ping Flood | 2.8ms | 6.7ms | 3.47ms | 8.8ms |
| UDP Flood | 3.1ms | 6.9ms | 4.1ms | 8.2ms |

The low response latency confirms the efficiency of the trust decay and threshold-based isolation mechanism of our system, which reacts based on consecutive trust violations before isolating a suspected node.

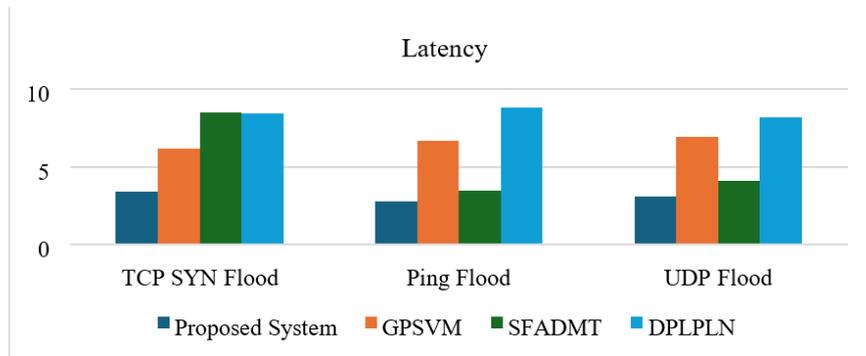

Figure 10. Latency

Network Throughput: We measured the volume of legitimate traffic successfully forwarded by the network before, during, and after attack detection. High throughput after detection indicates that the system mitigates the attack without disrupting normal traffic.

Table 5. Network Throughput

| Attack Type | Before Attack | During Attack | After Detection |
|---|---|---|---|
| TCP SYN Flood | 100 Mbps | 0 Mbps | 98 Mbps |
| Ping Flood | 100 Mbps | 0 Mbps | 96 Mbps |
| UDP Flood | 100 Mbps | 0 Mbps | 95 Mbps |

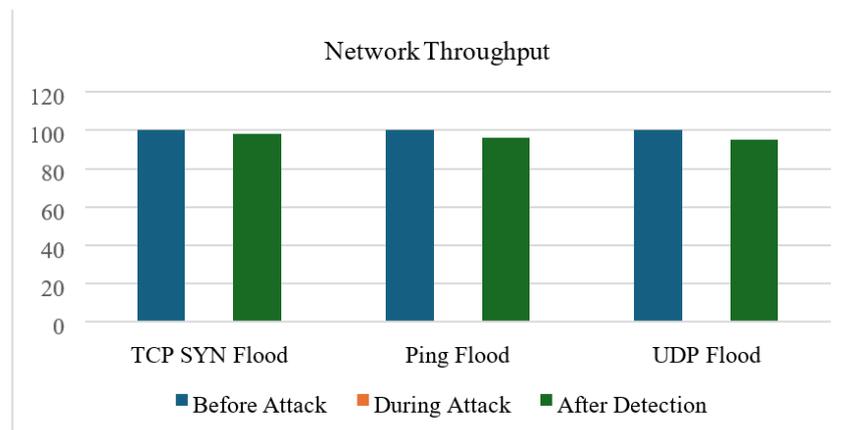

Figure 11. Network Throughput

Across all evaluated metrics, our hybrid trust-based system shows higher detection performance, lower false positives, faster response times, and better throughput compared to other existing solutions.

## 6. CONCLUSION

One promising approach toward achieving self-adaptive and resilient cybersecurity in Smart Internet of Things (SIoT) environments is the deployment of trust-based security mechanisms.

While existing studies have demonstrated the potential of such systems in controlled or simulated environments, real-world implementation still faces challenges, including scalability, interoperability, and responsiveness. This research advances the current body of knowledge by designing, implementing, and validating a lightweight trust-based intrusion detection and response system for SIoT networks. By combining knowledge-based and observational-based techniques, the mechanism provides a flexible, context-aware, and resilient method for trust computation. Trust update dynamics, propagation strategies, and threshold-based actions further improve security and performance. The system was evaluated using a combination of a SIoT testbed consisting of two smart cameras, four smart plugs, a smart lamp, a door sensor, and an alarm system. Three different attack types were carried out: TCP SYN, Ping flood and UDP flood attacks. The technique was further validated using OMNeT++ for scaling, and the approach consistently achieved high accuracy and responsiveness.

The results demonstrate that the system achieves a detection accuracy exceeding 98.3% under varying network loads, with improved response times compared to conventional anomaly-based DDoS detection methods. Additionally, the proposed trust-based approach significantly reduces false positives and enhances adaptive decision-making. The findings underscore that integrating trust evaluation and trust provenance tracking within IoT security architectures can substantially improve resilience, transparency, and energy efficiency. Finally, while our model performs well under the evaluated attack types, more complex multi-vector attacks and cross-domain trust inconsistencies were outside the study scope. Future work would focus on introducing Raspberry Pi 4 nodes, MQTT brokers, edge gateways, and a decentralized trust server to our IoT testbed, extending the architecture to enable cross-domain network security.

## ACKNOWLEDGMENT

The authors would like to acknowledge the support of Nottingham Trent University (NTU) for a fully funded studentship. The authors also declare that they have no known competing financial interests or personal relationships that could have appeared to influence the work reported in this paper.

## AUTHORS

**Oghenetejiri Okporokpo** is a PhD candidate of computer science at Nottingham Trent University UK.

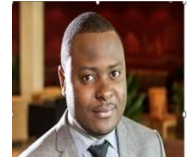

**Dr Funminiyi Olajide**, is a Senior Lecturer in Cyber Security and Forensics. He is part of the project supervision team.

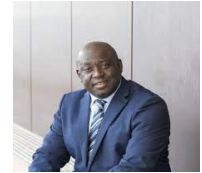

**Dr Nemitari Ajienka** is a Senior Lecturer at Nottingham Trent University, UK. He is part of the project supervision team.

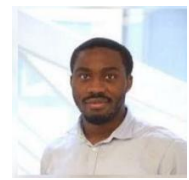

**Dr Xiaoqi Ma** is currently a Senior Lecturer in Department of Computer Science at Nottingham Trent University, UK. He is part of the project supervision team.

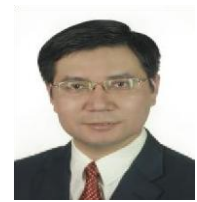